\documentclass[showpacs,preprintnumbers,amsmath,amssymb]{revtex4}
\usepackage{amsmath,amssymb,graphics,epsfig,subfigure}
\usepackage{color}

\begin{document}
\renewcommand{\baselinestretch}{1.3}

\title{The microstructure and Ruppeiner geometry of charged anti-de Sitter black holes in Gauss-Bonnet gravity: from the critical point to the triple point}

\author{Shao-Wen Wei \footnote{weishw@lzu.edu.cn},
  Yu-Xiao Liu \footnote{liuyx@lzu.edu.cn, corresponding author}}

\affiliation{Lanzhou Center for Theoretical Physics, Key Laboratory of Theoretical Physics of Gansu Province, School of Physical Science and Technology, Lanzhou University, Lanzhou 730000, People's Republic of China,\\
 Institute of Theoretical Physics $\&$ Research Center of Gravitation,
Lanzhou University, Lanzhou 730000, People's Republic of China,\\
 Academy of Plateau Science and Sustainability, Qinghai Normal University, Xining 810016, P. R. China}

\begin{abstract}
Ruppeiner geometry has been successfully applied in the study of the black hole microstructure by combining with the small-large black hole phase transition, and the potential interactions among the molecular-like constituent degrees of freedom are uncovered. In this paper, we will extend the study to the triple point, where three black hole phases coexist acting as a typical feature of black hole systems quite different from the small-large black hole phase transition. For the six-dimensional charged Gauss-Bonnet anti-de Sitter black hole, we thoroughly investigate the swallow tail behaviors of the Gibbs free energy and the equal area laws. After obtaining the black hole triple point in a complete parameter space, we exhibit its phase structures both in the pressure-temperature and temperature-horizon radius diagrams. Quite different from the liquid-vapor phase transition, a double peak behavior is present in the temperature-horizon radius phase diagram. Then we construct the Ruppeiner geometry and calculate the corresponding normalized curvature scalar. Near the triple point, we observe multiple negatively divergent behaviors. Positive curvature scalar is observed for the small black hole with high temperature, which indicates that the repulsive interaction dominates among the microstructure. Furthermore, we consider the variation of the curvature scalar along the coexisting intermediate and large black hole curves. Combining with the observation for different fluids, the result suggests that this black hole system behaves more like the argon or methane. Our study provides a first and preliminary step towards understanding black hole microstructure near the triple point, as well as uncovering the particular properties of the Gauss-Bonnet gravity.
\end{abstract}



\maketitle

\section{Introduction}

Motivated by the anti-de Sitter/conformal field theory (AdS/CFT) correspondence \cite{Maldacena,Gubser,Witten}, the study of the black hole thermodynamics and phase transition in AdS space has attracted a lot of attention. Quite different from the asymptotically flat spacetime, large black holes with positive heat capacity are thermodynamically stable. In particular, below a certain temperature, no black hole solution exits, while a pure thermal radiation is present. Hawking and Page \cite{Hawking1983b} discovered that a first-order phase transition between the pure radiation and stable large black hole emerges for this case, and now it is known as the Hawking-Page phase transition. Interestingly, such phase transition can be interpreted as the confinement/deconfinement phase transition of the gauge field \cite{Witten2}.

Further study has been extended to the charged and rotating black holes in AdS space. In the canonical ensemble, it was found that the black hole systems exhibit small-large black hole phase transition for fixed cosmological constant \cite{Chamblin,Chamblin2,Cognola}. Accompanied by the new interpretation of the cosmological constant that it plays a role of pressure \cite{Kastor}, the study of the black hole thermodynamics has been generalized to the extended phase space. One of the amazing results is that the precise analogy between the small-large black hole phase transition and the liquid-vapor phase transition of van der Waals (VdW) fluid has been completed by Mann and Kubiznak \cite{Kubiznak}. It was found that, similar to the VdW phase transition, the black hole phase transition point starts at zero, then increases with the temperature or pressure, and ends at a critical point, where the phase transition becomes a second-order one. Near the critical point, the universal exponents are calculated. The result indicates that they equal to that of the VdW fluid. This strengthens the observation that a black hole system behaves like an ordinary thermodynamic system. Subsequently, more interesting phase transitions and phase structures were observed \cite{Altamirano,AltamiranoKubiznak,Altamirano3,Dolan,Liu,Wei2,Frassino,
Cai,XuZhao,Hennigar,Hennigar2,Tjoa2,Ruihong}.

Since the phase transition is an indicator of the coexistence and competition of the dominated system components, a further study towards this is probing the microstructure of the black hole from the viewpoint of thermodynamics. This shall provide a complementary understanding on the black hole nature from the side of gravity. Motivated by this, we constructed a Ruppeiner geometry for the charged AdS black hole \cite{Weiw}. Via the number density, the microscopic and macroscopic physical quantities are closely linked with each other. After calculating the corresponding curvature scalar, it suggests that the microscopic interaction of the black hole can be reflected through the sign of the curvature scalar. The study deepens our understanding of the thermodynamics and microstructure of black holes. Further, we reconsidered this interesting approach. A more general Ruppeiner geometry approach was proposed in Refs. \cite{Weiwa2,WeiWeiWei}. For the charged AdS black hole, its heat capacity at constant volume usually vanishes, which spoils the Ruppeiner geometry. In order to uncover the information underlying in the geometry, we introduced a normalized curvature scalar \cite{Weiwa2}. Employing with this novel concept, we successfully examined the interaction dominating among the black hole microstructure. The result shows that repulsive interaction could dominate for the high temperature small black holes. Otherwise the attractive interaction dominates. This is significantly different from the VdW fluid, where the dominated repulsive interaction is always absent. Therefore, the characteristic behavior of the black hole microstructure is uncovered. In particular, the critical phenomena are observed for the normalized curvature scalar reflecting a universal property of the Ruppeiner geometry. Such study has also been extended to other black hole systems and more interesting results were obtained \cite{Dehyadegari,Zangeneh:2016fhy,Moumni,Miao,Du,Xuz,GhoshBhamidipati,Kumara2020,Yerra:2020oph,Wu:2020fij,
Vaid,Rizwan,Mansoori,Kumara,Mannw,Dehyadegariw,Wei2020d,YPhu,Xuzzm}.

On the other hand, Ruppeiner geometry \cite{Ruppeiner} has been applied to different fluid systems. It is worth to mentioning that in Ref. \cite{Ruppeinerpre}, by using the thermodynamic input from the NIST Chemistry WebBook, Ruppeiner evaluated the thermodynamic curvature scalar for fourteen pure fluids along their liquid-vapor coexistence curves varying from the critical point to the triple point. The scalars of six representative fluids were explicitly shown. Different features were observed. At the critical point, all these scalars have a negatively divergent behavior. While at the triple points, the scalars could be negative or positive. For hydrogen, helium, and oxygen, zero points of the scalars were found along the coexistence liquid curves. Significantly, the scalar has two zero points along the coexistence vapor curve for the water \cite{Ruppeinerpre}. Other interesting papers can also be found in Refs. \cite{Seftas,Castorina}.

These interesting studies provide us an excellent test for the black hole triple point, which is quite different from the small-large black hole phase transition. As a result, we wish to carry out the similar calculation for the black hole systems and to uncover the features of the scalar near the black hole triple points. Although the small-large black hole phase transition exists in most charged AdS black hole systems, the triple point is very limited. One of the black hole systems possessing the triple point is the charged Gauss-Bonnet (GB)-AdS black hole in six dimensions. In Ref. \cite{Liu}, we studied the phase transition in the charge-electric potential diagram. Interestingly, we found that there exist two non-monotonic behaviors on one isothermal curve for the six-dimensional case in a certain parameter region, which reveals the existence of a triple point. This result is confirmed by our following study in the extended phase space \cite{Wei2}, where the explicit triple point is shown. Also, this unique phenomenon in six dimensions was further confirmed in Ref. \cite{Frassino}.

Actually, the Ruppeiner geometry has been constructed for the black holes in the GB gravity. In Ref.
\cite{Wei2020a}, we first investigated the Ruppeiner geometry for the five-dimensional neutral GB-AdS black hole. The system admits an analytical coexistence curve, and gives us an opportunity to exactly study the phase transition and Ruppeiner geometry. Combining with the normalized curvature scalar, we uncovered an intriguing property for the black hole microstructure. When a system crosses the coexistence curve, its microstructure will undergo a sudden change. In general, the microscopic interaction changes accordingly. However, for the five-dimensional neutral GB-AdS black hole, the interaction keeps unchanged implied from its geometry. Even when the charge is present, this intriguing property still holds in the grand canonical ensemble \cite{Zhourun}. However, in other dimensions of spacetime, the case behaves quite differently. For the four-dimensional case, we found that the small-large black hole phase transition exists both in the canonical and grand canonical ensembles \cite{Wei2020d}. The corresponding geometry suggests that the repulsive interaction dominates for the high temperature small black holes, quite similar to the charged case in Einstein-Maxwell gravity. In higher dimensions more than five, we observed the similar phenomena \cite{WeiLL}. So in the canonical ensemble, the present of charge is the key to the repulsive interaction.

Although the Ruppeiner geometry has been well constructed for the black holes in GB gravity, the study near the triple points is still lacked. Therefore, the aim of this paper is to construct the Ruppeiner geometry for this case and examine the corresponding curvature scalar by varying from the critical point to the triple point. Comparing with the observations of different fluid systems \cite{Ruppeinerpre}, we are going to uncover the particular features near the triple point in the GB gravity for the first time.

The present work is organized as follows. In Sec. \ref{tobh}, we briefly review the thermodynamics for the six-dimensional charged GB-AdS black holes. In Sec. \ref{ptatp}, we study the phase transition of the black hole near the triple points. The features of the swallow tail behavior of the Gibbs free energy and the equal area law are examined near the triple point in details. After obtaining the triple point, we exhibit the phase structures in $P$-$T$ and $T$-$r_{\text{h}}$ diagrams. Further, we construct the Ruppeiner geometry in Sec. \ref{rgami}. The normalized curvature scalar is calculated. Microscopic interaction is reflected via the scalar. The characteristic behaviors of the scalar near the triple point are also investigated. In Sec. \ref{fcpttp}, we study the scalar along the coexisting intermediate and large black hole curve reminiscent of the liquid-vapor phase transition when the system varies from the critical point to the triple point. The results suggest this charged GB-AdS black hole system behaves like the argon or methane. Finally, we summarize and discuss our results in the last section.

\section{Thermodynamics of black holes}
\label{tobh}

Since the triple point we consider in this paper only exists in six dimensions in GB gravity, we take the dimension $d$=6.

The action of the GB gravity in six-dimensional spacetime is
\begin{eqnarray}
 S=\int d^{6}x\sqrt{-g}
 \Big[\frac{1}{16\pi G_{6}}(\mathcal{R}
    -2\Lambda
    +\alpha_{\text{GB}}\mathcal{L}_{\text{GB}})
  -\mathcal{L}_{\text{matter}}\Big], \label{action}
\end{eqnarray}
where $\Lambda$ is the cosmological constant and is negative in an AdS space. The parameter $\alpha_{\text{GB}}$ is the GB coupling constant of dimension $(\text{length})^{2}$. It can be regarded as the inverse string tension. The GB Lagrangian $\mathcal{L}_{\text{GB}}$ and the electromagnetic Lagrangian $\mathcal{L}_{\text{matter}}$ are in the following form
\begin{eqnarray}
&&\mathcal{L}_{\text{GB}}=\mathcal{R}_{\mu\nu\gamma\delta}\mathcal{R}^{\mu\nu\gamma\delta}
                    -4\mathcal{R}_{\mu\nu}\mathcal{R}^{\mu\nu}+\mathcal{R}^{2},\\
&&\mathcal{L}_{\text{matter}}=4\pi \mathcal{F}_{\mu\nu}\mathcal{F}^{\mu\nu}.
\end{eqnarray}
The Maxwell field strength is defined as
$\mathcal{F}_{\mu\nu}=\partial_{\mu}\mathcal{A}_{\nu}-\partial_{\nu}\mathcal{A}_{\mu}$
with $\mathcal{A}_{\mu}$ the vector potential.

A charged static, spherically symmetric black hole solution can be obtained by solving the field equation of the action (\ref{action}):
\begin{eqnarray}
 {\rm d}s^{2}=-f(r){\rm d}t^{2}+f^{-1}(r){\rm d}r^{2}+r^{2} {\rm d}\Omega^{2}_{4},\label{metric}
\end{eqnarray}
where ${\rm d}\Omega^{2}_{4}$ is the line element of the unit $S^4$ and the metric function is given by \cite{Boulware,Cai2,Wiltshire,Cvetic2}
\begin{eqnarray}
 f(r)=1+\frac{r^{2}}{2\alpha}
   \left(1-\sqrt{1+\frac{6\alpha M}{\pi r^5}-\frac{2 \alpha Q^2}{3 r^8}-\frac{16 \pi\alpha P}{5}}\right),
\end{eqnarray}
with $\alpha=6\alpha_{\text{GB}}$ and $P=-\frac{1}{8\pi}\Lambda$. The parameters $M$, $Q$, and $P$ are the black hole mass, charge, and pressure, respectively. Generally, the event horizon locates at the largest root of $f(r_{\text{h}})=0$, from which we can express the black hole mass in term of the horizon radius as
\begin{eqnarray}
 M=\frac{\pi  \left(6 \left(4 \pi  P r_{\text{h}}^8+5 \alpha
   r_{\text{h}}^4+5 r_{\text{h}}^6\right)+5 Q^2\right)}{45 r_{\text{h}}^3}.
\end{eqnarray}
The black hole temperature can be calculated by using the ``Euclidean trick"
\begin{eqnarray}
 T=\frac{f'(r_{\text{h}})}{4\pi}=
 \frac{8 \pi P r_{\text{h}}^8+2\alpha r_{\text{h}}^4+6
   r_{\text{h}}^6-Q^2}{16\pi\alpha r_{\text{h}}^5+8\pi r_{\text{h}}^7},\label{temper}
\end{eqnarray}
where the prime denotes the derivative with respect to $r_{\text{h}}$. The entropy, thermodynamic volume, and electric potential are, respectively, given by
\begin{eqnarray}
 S&=&\frac{2}{3} \pi ^2 r_{\text{h}}^4 \left(\frac{4 \alpha}{r_{\text{h}}^2}+1\right),\\
 V&=&\frac{8}{15} \pi ^2 r_{\text{h}}^5,\\
 \Phi&=&\frac{2 \pi  Q}{9 r_{\text{h}}^3}.
\end{eqnarray}
In Ref. \cite{Cai}, the GB coupling parameter $\alpha$ was also treated as a new thermodynamic quantity, and its conjugate quantity reads
\begin{eqnarray}
 \mathcal{A}=\frac{2 \pi  r_{\text{h}}}{3}-\frac{8}{3} \pi ^2 T r_{\text{h}}^2.
\end{eqnarray}
A simple calculation shows that these thermodynamic quantities satisfy the following first law
\begin{eqnarray}
 {\rm d} M=T{\rm d}S+\Phi {\rm d}Q+\mathcal{A}{\rm d}\alpha+V{\rm d}P.
\end{eqnarray}
Comparing with the conventional thermodynamic law, one can find that the black hole mass $M$ here plays the role of enthalpy of the system, so one can take $H\equiv M$. Moreover, the following Smarr law also holds
\begin{eqnarray}
 3H=4TS-2PV+2\mathcal{A}\alpha+3Q\Phi.
\end{eqnarray}
The local thermodynamic stability is determined by the heat capacity of the system. Positive or negative value implies stable or unstable of the black hole. For this charged GB-AdS black hole, the heat capacity can be calculated as
\begin{eqnarray}
 C_{P, Q, \alpha}&=&T\left(\frac{\partial S}{\partial T}\right)_{P, Q, \alpha}\nonumber\\
 &=&\frac{64 \pi ^3 r_{\text{h}}^7 \left(2 \alpha
   +r_{\text{h}}^2\right)^3T}
   {3 \left[
    8 \pi  P r_{\text{h}}^{10}
   +6  (8 \pi  \alpha P-1)r_{\text{h}}^8
   +6 \alpha    r_{\text{h}}^6
   -4 \alpha ^2 r_{\text{h}}^4
    + 7 Q^2 r_{\text{h}}^2 +10 Q^2 \alpha
   \right]}.
\end{eqnarray}
Obviously, the heat capacity can be positive or negative due to the different values of these parameters. For the extremal black hole with $T$=0, one then obtains a vanishing heat capacity. On the contrary, the global stability is determined by the Gibbs free energy
\begin{eqnarray}
 G=H-TS=\frac{\pi  \left(5 Q^2 \left(20 \alpha +7
   r_{\text{h}}^2\right)-6 r_{\text{h}}^4 \left(-20 \alpha ^2+r_{\text{h}}^4
   (48 \pi  \alpha  P-5)+4 \pi  P r_{\text{h}}^6+5 \alpha
   r_{\text{h}}^2\right)\right)}{180 r_{\text{h}}^3 \left(2 \alpha
   +r_{\text{h}}^2\right)}.
\end{eqnarray}
As we know, thermodynamic system always prefers a phase of the lowest free energy. The phase has a higher Gibbs free energy might be metastable or unstable if its heat capacity is positive or negative.

In what follows, we would like to rescale these thermodynamic quantities by the black hole charge, for examples
\begin{eqnarray}
 \tilde{T}=TQ^{\frac{1}{3}},\quad
 \tilde{P}=PQ^{\frac{2}{3}},\quad
 \tilde{r}_{\text{h}}=r_{\text{h}}Q^{-\frac{1}{3}},\quad
 \tilde{V}=VQ^{-\frac{5}{3}},\quad
 \tilde{\alpha}=\alpha Q^{-\frac{1}{2}},\quad
 \tilde{G}=G Q^{-1}.
\end{eqnarray}
Or equivalently, we set $Q$=1 for simplicity.

\section{Phase transitions and triple points}
\label{ptatp}

As shown in Refs. \cite{Liu,Wei2}, in GB gravity a triple point exists and only exists in six dimensions. This result was also confirmed in Ref. \cite{Frassino}. In this section, we would like to examine such phenomenon in a complete parameter space. New phase structure will also be disclosed for the first time.

\subsection{Swallow tail behaviors of Gibbs free energy}

The small-large black hole phase transition is a first-order one, during which the Gibbs free energy keeps unchanged. Intuitively, the phase transition point is determined by the swallow tail behavior of the Gibbs free energy. In what follows, we will show the interesting swallow tail behaviors when the triple point is present.

%%%%%%%%%%%%%%%%%%%%%%%%%%%
\begin{figure}
\center{\subfigure[]{\label{GibssSkip}
\includegraphics[width=7cm]{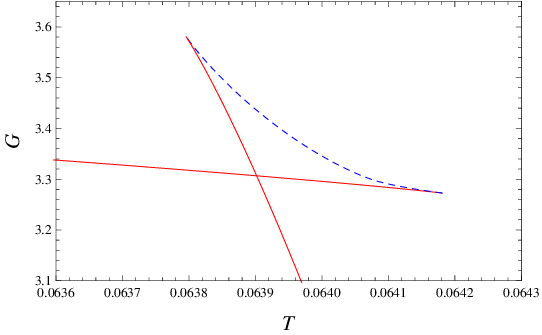}}
\subfigure[]{\label{Gibssgb}
\includegraphics[width=7cm]{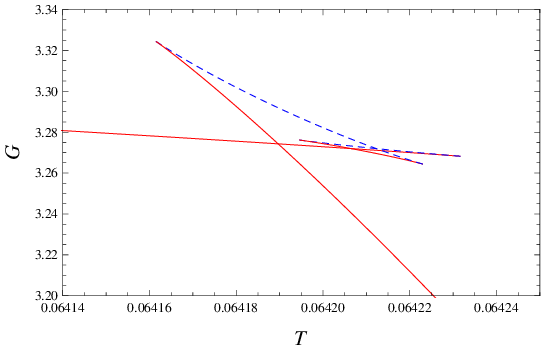}}
\subfigure[]{\label{Gibssgc}
\includegraphics[width=7cm]{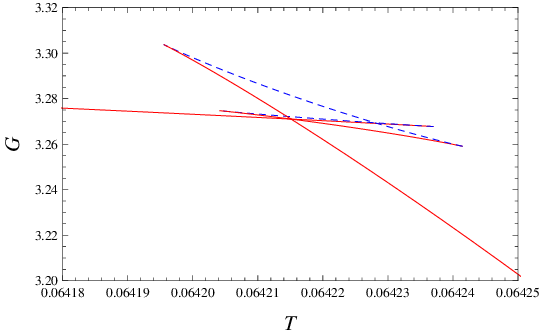}}
\subfigure[]{\label{Gibssgd}
\includegraphics[width=7cm]{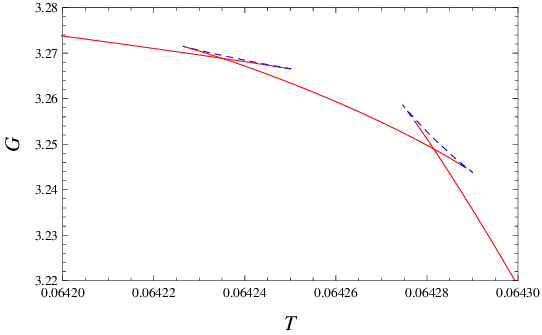}}
\subfigure[]{\label{Gibssge}
\includegraphics[width=7cm]{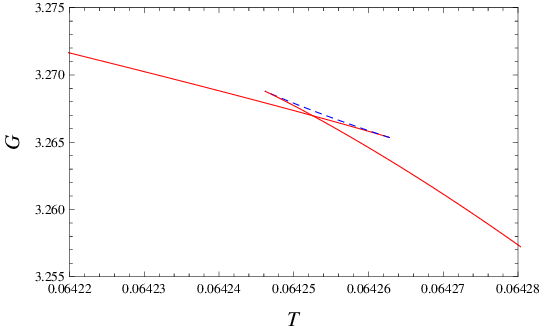}}
\subfigure[]{\label{Gibssgf}
\includegraphics[width=7cm]{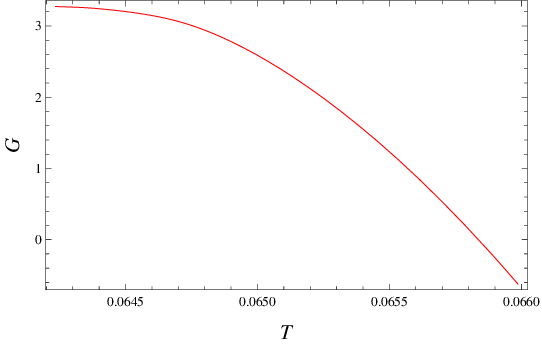}}}
\caption{Behaviors of the Gibbs free energy as a function of the temperature with $\alpha$=3.05. These branches described by the red solid curves and blue dashed cures are local stable (positive heat capacity) or unstable (negative heat capacity), respectively. (a) $P$=0.0062. (b) $P$=0.00637. (c) $P$=$P_{\text{t}}$=0.006387. (d) $P$=0.00643. (e) $P$=0.00647. (f) $P$=0.0067.}\label{ppGibssgf}
\end{figure}
%%%%%%%%%%%%%%%%%%%%%%%%%%%

We describe the behaviors of the Gibbs free energy in Fig. \ref{ppGibssgf} for different values of the pressure with $\alpha$=3.05. The red solid curves have positive heat capacity, which means the corresponding black holes are local stable. While the black holes correspond to the blue dashed curves with negative heat capacity are local unstable. For $P$=0.0062 shown in Fig. \ref{GibssSkip}, there displays a swallow tail behavior indicating a first-order phase transition. Considering that a system always prefers a phase of the lowest Gibbs free energy, so it prefers a small black hole at low temperature and a large black hole at high temperature. The phase transition just occurs at the intersection point of these two stable small and large black hole branches. Thus, with the increase of the temperature, the system goes along the small black hole branch at first, and then transits to the large black hole branch. Increasing the pressure such that $P$=0.00637 (see Fig. \ref{Gibssgb}), a stable intermediate black hole branch emerges from the unstable black hole branch. Two swallow tail behaviors are observed. Nevertheless, this new branch has a higher Gibbs free energy, and it will not influence the black hole phase structure. Therefore, there is still only one small-large black hole phase transition. When we take $P$=$P_{\text{t}}$=0.006387, an interesting behavior is displayed in Fig. \ref{Gibssgc}. All these three stable black hole branches intersect at the same one point, which means that the corresponding small, intermediate, and large black holes have the same Gibbs free energy. Hence, these black holes can coexist at that point, which is just the triple point. When $P$=0.00643, we clearly observe a double swallow tail behavior in Fig. \ref{Gibssgd}. These three stable black hole branches intersect at two different points, which denote the small-intermediate black holes and the intermediate-large black hole phase transition, respectively. So with the increase of the temperature, the system goes along small, intermediate, large black hole branches, respectively. For such isobaric curve, the system will undergo two phase transitions at two different temperatures. Further when setting $P$=0.00647, we find that the double swallow tail behavior disappears, while only one swallow tail behavior is present, which indicates that only a liquid-vapor type phase transition exists. For a high pressure $P$=0.0067, see Fig. \ref{Gibssgf}, all the swallow tail behaviors disappear. Only one stable black hole branch exists, which implies that we cannot distinguish the small, intermediate, and large black holes any more. And thus there is no phase transition for this case.

In a word, from above behaviors of the Gibbs free energy, we observe that there exists the triple point. Actually, this phenomenon only occurs in six-dimensional GB gravity.

\subsection{Equal area laws}

Generally, the phase transition point is determined by the swallow tail behavior. Alternatively, it can also be obtained by the Maxwell equal area law.

At the intersection point of the swallow tail behavior, the change of the Gibbs free energy of these two coexistence phases vanishes
\begin{eqnarray}
 {\rm d} G=0.
\end{eqnarray}
By using the differential form of the Gibbs free energy ${\rm d}G=-S{\rm d}T+\Phi {\rm d}Q+\mathcal{A}{\rm d}\alpha+V{\rm d}P$, we have
\begin{eqnarray}
 V{\rm d} P=0,
\end{eqnarray}
for fixed $T$, $Q$, and $\alpha$. Considering a phase transition between these two coexistence phases at fixed temperature $T^*$ and pressure $P^*$, one  has
\begin{eqnarray}
 \oint V(Q, \alpha, T^*, P){\rm d} P=0.
\end{eqnarray}
This formula can also be expressed as
\begin{eqnarray}
 \int_{V_{\text{s}}}^{V_{\text{l}}} P(Q, \alpha, T^*, V){\rm d} V=P^*(V_{\text{l}}-V_{\text{s}}),\label{eqmax}
\end{eqnarray}
where the indexes ``s" and ``l" stand for the coexisting small and large black holes, respectively. Solving (\ref{temper}), we obtain the equation of state,
\begin{eqnarray}
 P=\frac{8\pi^{12/5}-45\times(60V^6)^{1/5}+30T\times(60^2\pi^3V^7)^{1/5}-\alpha(60\pi V)^{4/5}+240\pi^{7/5}\alpha TV}{30(2\times15^3\pi V^8)^{1/5}}.   \label{equat}
\end{eqnarray}
Note that the black hole charge has been set to one. Substituting it into (\ref{eqmax}), the Maxwell equal area law can be expressed as
\begin{eqnarray}
 \frac{5
   \left(r_{\text{hl}}^3-r_{\text{hs}}^3\right)}{6 r_{\text{hl}}^3
   r_{\text{hs}}^3}=4\pi P^* ( r_{\text{hl}}^5-r_{\text{hs}}^5)-5 \pi T^* (r_{\text{hl}}^4- r_{\text{hs}}^4)+5(r_{\text{hl}}^3-r_{\text{hs}}^3)-5 \alpha\left(r_{\text{hl}}-r_{\text{hs}}\right) \left(4 \pi T^*\left(r_{\text{hl}}+r_{\text{hs}}\right)-1\right),
\end{eqnarray}
where we have replaced the thermodynamic volume with the horizon radius for simplicity. In addition, the coexisting small and large black holes also satisfy the equation of state (\ref{equat}), which gives
\begin{eqnarray}
 2 r_{\text{hs}}^4 \left(\alpha +4 \pi P^*r_{\text{hs}}^4-8\pi
   \alpha T^*r_{\text{hs}}-4\pi T^* r_{\text{hs}}^3+3 r_{\text{hs}}^2\right)-1=0,\\
 2 r_{\text{hl}}^4 \left(\alpha +4 \pi P^*r_{\text{hl}}^4-8\pi
   \alpha T^*r_{\text{hl}}-4\pi T^* r_{\text{hl}}^3+3 r_{\text{hl}}^2\right)-1=0.
\end{eqnarray}
For given $\alpha$ and $T^*$, we can obtain $P^*$, $V_{\text{s}}$ and $V_{\text{l}}$, respectively, by solving the above equations.

%%%%%%%%%%%%%%%%%%%%%%%%%%%
\begin{figure}
\center{\subfigure[]{\label{Equalarealawa}
\includegraphics[width=7cm]{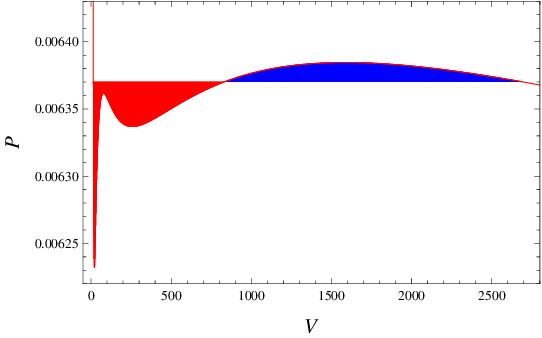}}
\subfigure[]{\label{Equalarealawb}
\includegraphics[width=7cm]{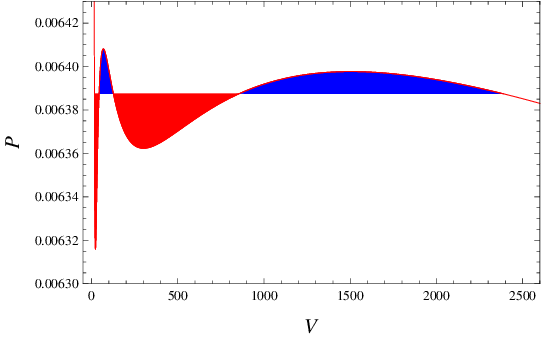}}
\subfigure[]{\label{Equalarealawc}
\includegraphics[width=7cm]{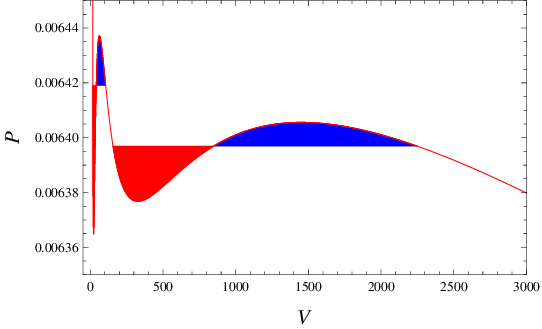}}
\subfigure[]{\label{Equalarealawd}
\includegraphics[width=7cm]{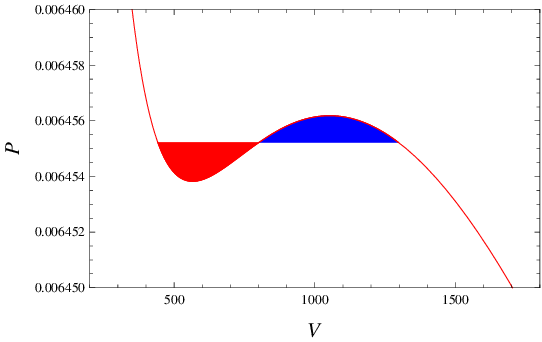}}}
\caption{The equal area law in the $P$-$V$ plane with $\alpha$=3.05. (a) $T$=0.06419. (b) $T$=$T_{\text{t}}$=0.064215. (c) $T$=0.06423. (d) $T$=0.06432.}\label{ppEqualarealawd}
\end{figure}
%%%%%%%%%%%%%%%%%%%%%%%%%%%

In order to clearly exhibit the equal area law for the GB black hole near the triple point, we plot the isothermal curves with $\alpha$=3.05 in Fig. \ref{ppEqualarealawd}. Noting that the equal area law only holds for the thermodynamic volume $V$ rather than the horizon radius $r_{\text{h}}$ \cite{Wei3}, we plot it in the $P$-$V$ plane. The temperature is set to $T$=0.06419, 0.064215, 0.06423, and 0.06423. Obviously, the case described in Fig. \ref{Equalarealawd} is the conventional behavior, which is because that the temperature is high enough, and only one small-large or small-intermediate black hole phase transition exists. One can construct two equal areas (see the shadow areas marked with the red and blue colors) along the isothermal curve. When the temperature $T$=0.06419 is smaller than that of the triple point, we describe the behavior in Fig. \ref{Equalarealawa}. The isothermal curve has two local minimal points, and is different from the case shown in Fig. \ref{Equalarealawd}. These two equal areas can also be constructed, which is still like Fig. \ref{Equalarealawd}. If the temperature takes the value of the triple point, the behavior becomes interesting, see Fig. \ref{Equalarealawb}. It is easy to see that two pair equal areas appear. While the values of these two pair equal area are not necessarily equal. For our case, the first two areas equal 0.001044, and the left two equal 0.010348. Slightly increasing the temperature, we can still construct two pair equal areas, see Fig \ref{Equalarealawc}. However, these two pair areas are not equal to each other, and the corresponding phase transition pressures do not coincide, which is different from the case of the triple point.

In summary, in this subsection, we construct several different cases of the equal area behaviors. Significantly, at the triple point, we have two pair equal areas constructed by one isobaric horizontal line and the isothermal curve. This is the characteristic behavior of the equal area law for the triple point.

\subsection{Triple points}

As shown above, there exists the triple point for the six-dimensional charged GB AdS black hole. So the problem that how to determine the triple point emerges. Now we attempt to develop an effective approach.

Since there are several parameters, it is hard to scan the parameter space to find the triple point. So the first thing is to shrink the possible parameter space for the triple point.

\begin{figure}
\includegraphics[width=7cm]{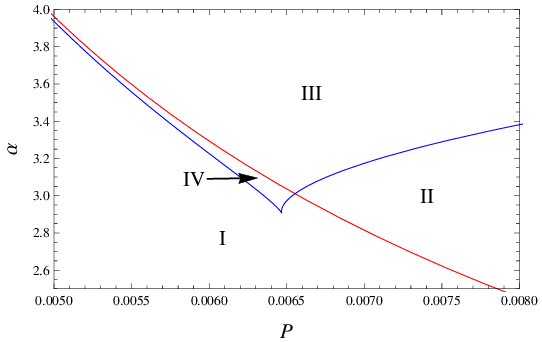}
\caption{Critical point in the $\alpha$-$P$ plane. The parameter space is divided into four parts, which have different numbers of the turning points of the temperature.}\label{AlphaT}
\end{figure}

As noted in our previous paper \cite{Wei2} or from the behaviors of the Gibbs free energy in Fig. \ref{ppGibssgf}, we find that the triple point exists at the case where we have four turning points at each isobaric line. For the purpose, we solve the critical point from $(\partial_{r_{\text{h}}}T)_{P, \alpha}=(\partial_{r_{\text{h}}, r_{\text{h}}}T)_{P, \alpha}=0$. It is quite hard to express the pressure $P$ and horizon radius $r_{\text{h}}$ in terms of $\alpha$. However, we can obtain the following results
\begin{eqnarray}
 P_{1,2}&=&\frac{20+3 r_{\text{h}}^6\pm\sqrt{3} \sqrt{44 r_{\text{h}}^6+75}}{24\pi r_{\text{h}}^8},\\
 \alpha_{1,2}&=&\frac{15+3 r_{\text{h}}^6\mp\sqrt{3} \sqrt{44 r_{\text{h}}^6+75}}{6 r_{\text{h}}^4}.
\end{eqnarray}
In order to clearly show the critical point, we plot it in the $\alpha$-$P$ plane in  Fig. \ref{AlphaT} by varying $r_{\text{h}}$ freely. We observe that for fixed $\alpha$, there may be one, two, or three critical points. However, one should keep in mind that not all of these critical points are stable. Some of them are excluded due to that they have a higher Gibbs free energy than other black hole branches. These critical point curves divide the parameter space into four regions. The calculation shows that there will be two, zero, two, and four turning points of the temperature for regions I, II, III, and IV, respectively. This suggests that the triple point is only present in region IV. Therefore, after this treatment, the region that the triple point exists is greatly shrunk.

%%%%%%%%%%%%%%%%%%%%%%%%%%%
%%%%%%%%%%%%%%%%%%%%%%%%%%%

Next, we will search for the triple point in region IV. In the last subsection, we show that the triple point can be determined by the equal area law shown in Fig. \ref{Equalarealawb}. Here we denote these three phases as the small, intermediate, and large black holes with radiuses $r_{\text{hs}}$, $r_{\text{hi}}$, and $r_{\text{hl}}$, respectively. The first condition is that all the three coexistence phases satisfy the equation of state (\ref{equat}), and thus, we have
\begin{eqnarray}
 2 r_{\text{hs}}^4 \left(\alpha +4 \pi P_{\text{t}}r_{\text{hs}}^4-8\pi
   \alpha T_{\text{t}}r_{\text{hs}}-4\pi T_{\text{t}} r_{\text{hs}}^3+3 r_{\text{hs}}^2\right)-1=0,\\
 2 r_{\text{hi}}^4 \left(\alpha +4 \pi P_{\text{t}}r_{\text{hi}}^4-8\pi
   \alpha T_{\text{t}}r_{\text{hi}}-4\pi T_{\text{t}} r_{\text{hi}}^3+3 r_{\text{hi}}^2\right)-1=0,\\
 2 r_{\text{hl}}^4 \left(\alpha +4 \pi P_{\text{t}}r_{\text{hl}}^4-8\pi
   \alpha T_{\text{t}}r_{\text{hl}}-4\pi T_{\text{t}} r_{\text{hl}}^3+3 r_{\text{hl}}^2\right)-1=0.
\end{eqnarray}
$P_{\text{t}}$ and $T_{\text{t}}$ stand for the pressure and temperature of the triple point. In addition, we have two more equations from the equal area law,
\begin{eqnarray}
 \frac{5
   \left(r_{\text{hi}}^3-r_{\text{hs}}^3\right)}{6 r_{\text{hi}}^3
   r_{\text{hs}}^3}
   &=&4\pi P_{\text{t}} ( r_{\text{hi}}^5-r_{\text{hs}}^5)-5 \pi T_{\text{t}} (r_{\text{hi}}^4- r_{\text{hs}}^4)+5(r_{\text{hi}}^3-r_{\text{hs}}^3)-5 \alpha\left(r_{\text{hi}}-r_{\text{hs}}\right) \left(4 \pi T_{\text{t}}\left(r_{\text{hi}}+r_{\text{hs}}\right)-1\right),\\
 \frac{5
   \left(r_{\text{hl}}^3-r_{\text{hi}}^3\right)}{6 r_{\text{hl}}^3
   r_{\text{hi}}^3}
   &=&4\pi P_{\text{t}} ( r_{\text{hl}}^5-r_{\text{hi}}^5)-5 \pi T_{\text{t}} (r_{\text{hl}}^4- r_{\text{hi}}^4)+5(r_{\text{hl}}^3-r_{\text{hi}}^3)-5 \alpha\left(r_{\text{hl}}-r_{\text{hi}}\right) \left(4 \pi T_{\text{t}}\left(r_{\text{hl}}+r_{\text{hi}}\right)-1\right).
\end{eqnarray}
Now we can solve these five equations to search for the triple point in region IV. For a given $\alpha$, we have a set of values of $P_{\text{t}}$, $T_{\text{t}}$, $r_{\text{hs}}$, $r_{\text{hi}}$, and $r_{\text{hl}}$. Eventually, we obtain a curve for the triple point by varying the GB coupling parameter $\alpha$.

\begin{figure}
\center{\subfigure[]{\label{PTtrip}
\includegraphics[width=5cm]{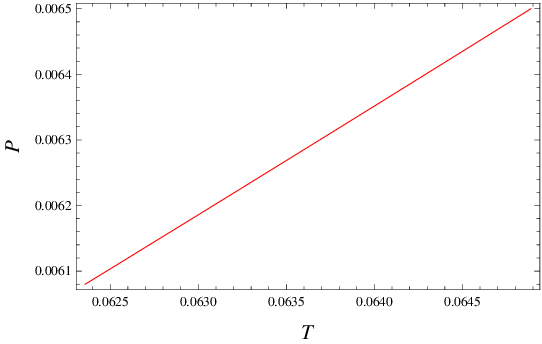}}
\subfigure[]{\label{Talphatrip}
\includegraphics[width=5cm]{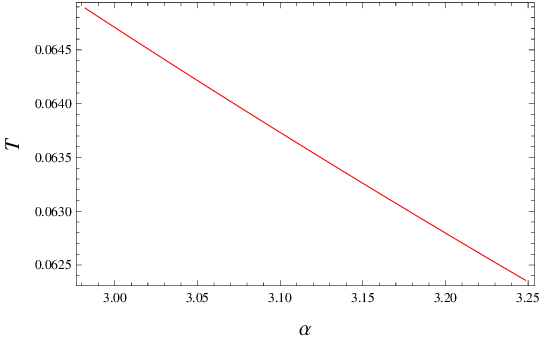}}
\subfigure[]{\label{Palpha}
\includegraphics[width=5cm]{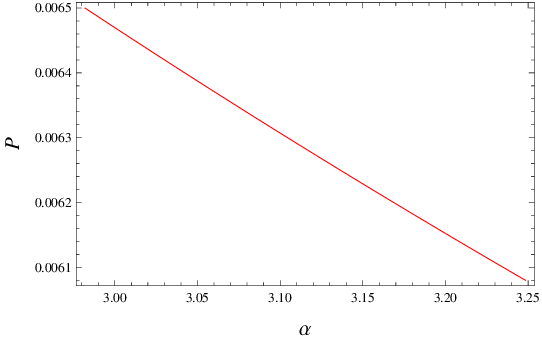}}}
\caption{Triple points in different planes. (a) The $P$-$T$ plane. (b) The $T$-$\alpha$ plane. (c) The $P$-$\alpha$ plane.}\label{ppPalpha}
\end{figure}

Here we depict the triple point in the $P$-$T$, $T$-$\alpha$, $P$-$\alpha$ planes in Fig. \ref{ppPalpha}. In Fig. \ref{PTtrip}, we observe that the pressure of the triple point seems to linearly increase with the temperature by varying $\alpha$. At the same time, we can find that with the increase of $\alpha$, both the temperature and pressure decrease linearly, for detail see Figs. \ref{Talphatrip} and \ref{Palpha}. Moreover, we also list the corresponding values of the triple points in Table \ref{tab1}, which might be useful for further studying the triple point.

%%%%%%%%%%%%%%%%%%%%%%%%%%%
%%%%%%%%%%%%%%%%%%%%%%%%%%%

%%%%%%%%%%%%%%%%%%%%%%%%%%%%%
\begin{table}[h]
\begin{center}
\begin{tabular}{cccccc}
  \hline\hline
   $\alpha$ & $P_{\text{t}}$   & $T_{\text{t}}$ &  $r_{\text{hs}}$ & $r_{\text{hi}}$ & $r_{\text{hl}}$ \\\hline
 3.00 & 0.006470 & 0.064709 & 1.35827 & 1.69610 & 3.45753 \\
 3.05 & 0.006387 & 0.064215 & 1.25552 & 1.88801 & 3.39893 \\
 3.10 & 0.006307 & 0.063732 & 1.19447 & 2.04870 & 3.32974 \\
 3.15 & 0.006229 & 0.063260 & 1.14928 & 2.20987 & 3.24410 \\
 3.20 & 0.006152 & 0.062797 & 1.11300 & 2.39329 & 3.12722 \\ \hline\hline
\end{tabular}
\caption{The corresponding values for the triple points.}\label{tab1}
\end{center}
\end{table}
%%%%%%%%%%%%%%%%%%%%%%%%%%%%%%%%

\subsection{Phase diagrams}

If there is a triple point, the phase diagram or structure will be different from that of the VdW fluid. Here, we would like to examine them for the black hole system.

\begin{figure}
\center{\subfigure[]{\label{PHSTRPT}
\includegraphics[width=7cm]{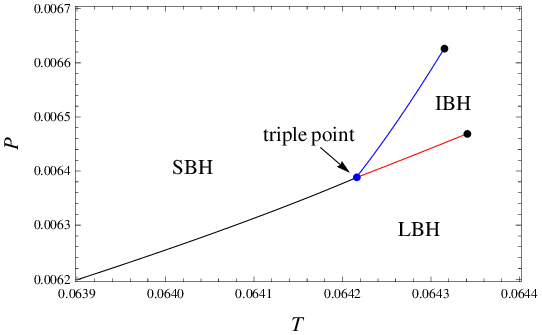}}
\subfigure[]{\label{PHSTRTRh}
\includegraphics[width=7cm]{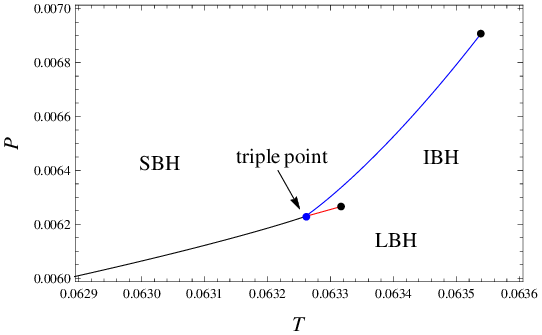}}}
\caption{Phase structure near the triple point for the six-dimensional charged GB-AdS black hole in the $P$-$T$ diagram. Blue and black dots denote the triple points and critical points, respectively. The abbreviations ``SBH", ``IBH", and ``LBH" are for the small, intermediate and large black holes, respectively. (a) $\alpha$=3.05. (b) $\alpha$=3.15.}\label{ppPHSTRTRh}
\end{figure}

Taking $\alpha$=3.05 and 3.15, we describe the phase structure in the $P$-$T$ plane in Fig. \ref{ppPHSTRTRh}, which is the same as that given in Ref. \cite{Wei2}. The small, intermediate, and large black holes are clearly exhibited. The blue dot denotes the triple point, where these small, intermediate, and large black hole phases coexist. While the black dots denote two critical points with different temperatures and pressures. The black, blue, and red curves stand for the coexistence curves of the small and large black holes, the small and intermediate black holes, and the intermediate and large black holes. Comparing with Figs. \ref{PHSTRPT} and \ref{PHSTRTRh}, we find that the lower critical point gets more close to the triple point for a bigger $\alpha$.

%%%%%%%%%%%%%%%%%%%%%%%%%%%
%%%%%%%%%%%%%%%%%%%%%%%%%%%

%%%%%%%%%%%%%%%%%%%%%%%%%%%
\begin{figure}
\center{\subfigure[]{\label{PHStrh305a}
\includegraphics[width=7cm]{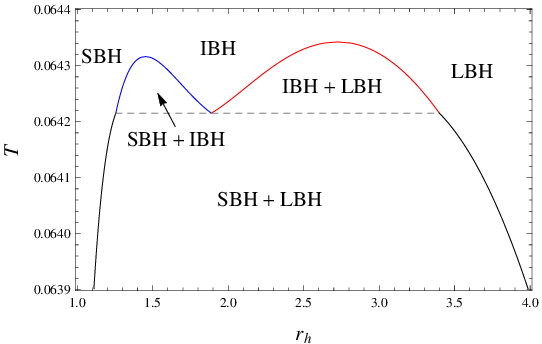}}
\subfigure[]{\label{PHStrh315b}
\includegraphics[width=7cm]{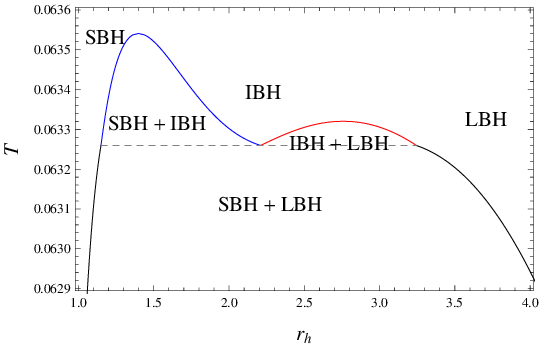}}}
\caption{Phase structure near the triple point for the six-dimensional charged GB-AdS black hole in the $T$-$r_{\text{h}}$ diagram. The dashed horizontal lines stand for the triple point, and these two peaks are for the critical points. (a) $\alpha$=3.05. (b) $\alpha$=3.15.}\label{ppPHStrh315b}
\end{figure}
%%%%%%%%%%%%%%%%%%%%%%%%%%%

On the other hand, we would like to show the phase structure in the $T$-$r_{\text{h}}$ diagram in Fig. \ref{ppPHStrh315b}. Interestingly, the double peak pattern is present. The similar picture has been reported in our previous study \cite{WeiWang}. Three coexistence regions and three black hole phase regions are clearly displayed. It is worth to point out that the triple point now is described by dashed horizontal lines, where the small, intermediate, and large black holes coexist. So the degeneracy of the coexistence regions and the triple points shown in Fig. \ref{ppPHSTRTRh} is eliminated in the $T$-$r_{\text{h}}$ diagram. Moreover, for small $\alpha$, the left peak is lower than the right one. Such case reverses for big $\alpha$.

\section{Ruppeiner geometry and microscopic interactions}
\label{rgami}

In this section, we would like to construct the Ruppeiner geometry for the charged GB-AdS black hole, and then discuss the interactions near the triple point via calculating the corresponding curvature scalar.

\subsection{Ruppeiner geometry}

Starting from the Boltzmann entropy formula and expanding the entropy $S$ near its local maximum $x^{\mu}_0$ ($\mu$=1,2), one can arrive for a two-parameter thermodynamic system \cite{Ruppeiner}
\begin{equation}
 S=S(x^{\mu}_0)+\left. \frac{1}{2}\frac{\partial^{2}S}{\partial x^{\mu} \partial x^{\nu}}
              \right|_{x^{\mu}_0} \Delta x^{\mu} \Delta x^{\nu},
\end{equation}
where we have neglected the influence of the environment due to its vast entropy. One should note that the fluctuating parameters $x^{1}$ and $x^{2}$ are additive. Then the probability of finding the system in the internals $x_0\sim (x_0 + \Delta x_0)$ and $x_1\sim (x_1 + \Delta x_1)$ can be expressed as
\begin{eqnarray}
 P(x^{0},x^{1}) \propto e^{-\frac{1}{2}\Delta l^{2}},
\end{eqnarray}
where
\begin{eqnarray}
  \Delta l^{2}&=&-\frac{1}{k_{\rm B}}g_{\mu\nu} \Delta x^{\mu}\Delta x^{\nu},\label{Ds}\\
 g_{\mu\nu}&=& \left.\frac{\partial^{2}S}{\partial x^{\mu}\partial x^{\nu}} \right|_{x^{\mu}_0} .\label{gmunu}
\end{eqnarray}
Here $\Delta l^{2}$ is known as the line element of the Ruppeiner geometry. For a given thermodynamic system, one is allowed to calculate the curvature scalar from the equation of state. This has also been applied to different fluid systems. On the other hand, the line element (\ref{Ds}) can also been cast into a more convenient form
\begin{equation}
 {\rm d} l^{2}=\frac{C_{V}}{T^{2}}{\rm d} T^{2}-\frac{(\partial_{V}P)_{T}}{T}{\rm d} V^{2},\label{xxy}
\end{equation}
where we adopt the internal energy $U$ and thermodynamic volume $V$ as the original fluctuating coordinates, and then change to $T$ and $V$. Obviously, if one knows the equation of state $P=P(T, V)$, the curvature scalar will be easily calculated following the Riemann approach. For a charged AdS black hole, we usually have $C_{V}$=0, which will spoil the geometry. In order to expose the result hiding behind the vanishing $C_{V}$, we assume that it keeps an infinitesimal constant, and introduce a normalized curvature scalar \cite{Weiwa2,WeiWeiWei}
\begin{equation}
 R_{\rm N}=R*C_{V}=\frac{1}{2}-\frac{T^{2}(\partial_{V,T}P)^{2}-2T^{2}(\partial_{V}P)(\partial_{V,T,T}P)}{2(\partial_{V}P)^{2}}.
 \label{scalarcurv}
\end{equation}
Furthermore, if $P$ is linear with $T$, it can be simplified as
\begin{equation}
 R_{\rm N}=R*C_{V}=\frac{1}{2}\left(1-\frac{T^{2}(\partial_{V,T}P)^{2}}{(\partial_{V}P)^{2}}\right).
 \label{scalarabc}
\end{equation}
This new concept has been successfully applied to many black hole systems and interesting black hole microscopic properties are observed.

\subsection{Normalized curvature scalar}

\begin{figure}
\center{\includegraphics[width=7cm]{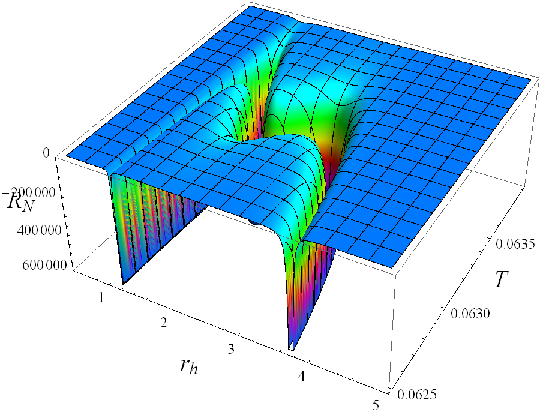}}
\caption{Behavior of the curvature scalar $R_{\rm N}$ for $\alpha$=3.15.}\label{RN3D}
\end{figure}

Now we turn to calculate the curvature scalar for the black hole system. By making use of the equation of state (\ref{equat}), we obtain the curvature scalar via (\ref{scalarabc}), which can be expressed in terms of $T$ and $V$ as
\begin{equation}
 R_{\rm N}=-\frac{\left(2 \alpha  r_{\text{h}}^4+3 r_{\text{h}}^6-2\right)
   \left(2 \alpha  r_{\text{h}}^4 \left(12 \pi  T
   r_{\text{h}}-1\right)+r_{\text{h}}^6 \left(4 \pi  T
   r_{\text{h}}-3\right)+2\right)}{2 \left(2 \alpha  r_{\text{h}}^4
   \left(6 \pi  T r_{\text{h}}-1\right)+r_{\text{h}}^6 \left(2 \pi  T
   r_{\text{h}}-3\right)+2\right)^2}.\label{rrrn}
\end{equation}
In order to show the behavior of $R_{\rm N}$ near the triple point, we describe it in Fig. \ref{RN3D} for different $T$ and $r_{\text{h}}$. Quite different from the VdW like phase transition, $R_{\rm N}$ here exhibits a more interesting phenomenon. Of particular interest is that, near the triple point, there may be four negative divergent points. A double peak pattern is present just analogy to the phase structures shown in Fig. \ref{ppPHStrh315b}.

%%%%%%%%%%%%%%%%%%%%%%%%%%%
%%%%%%%%%%%%%%%%%%%%%%%%%%%

%%%%%%%%%%%%%%%%%%%%%%%%%%%
\begin{figure}
\center{\subfigure[]{\label{Rnrhtta}
\includegraphics[width=7cm]{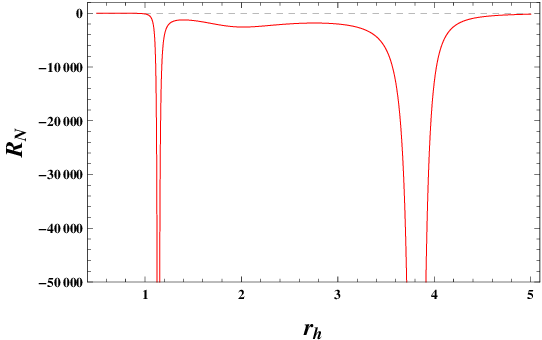}}
\subfigure[]{\label{Rnrht0629b}
\includegraphics[width=7cm]{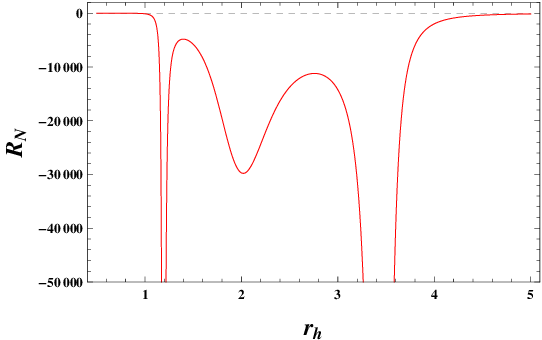}}
\subfigure[]{\label{Rnrhtc1c}
\includegraphics[width=7cm]{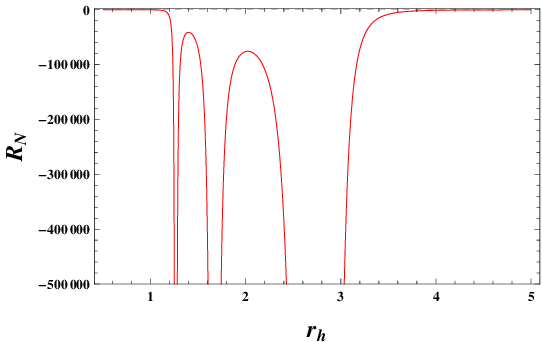}}
\subfigure[]{\label{Rnrht0634d}
\includegraphics[width=7cm]{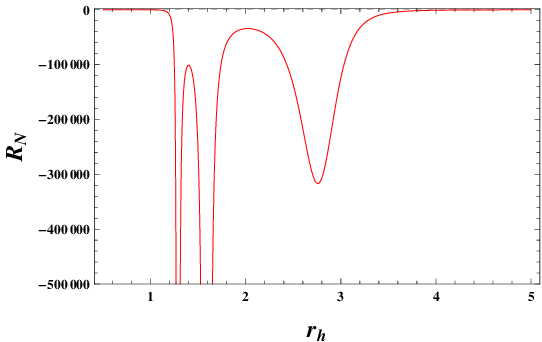}}
\subfigure[]{\label{Rnrhttc2e}
\includegraphics[width=7cm]{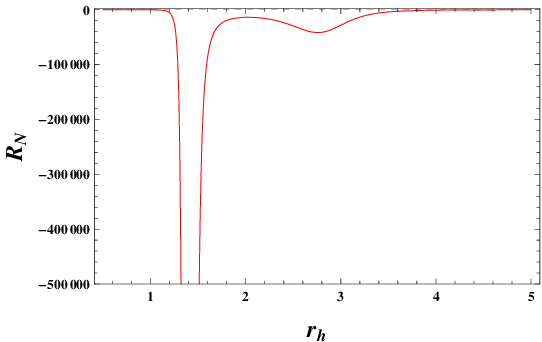}}
\subfigure[]{\label{Rnrht064f}
\includegraphics[width=7cm]{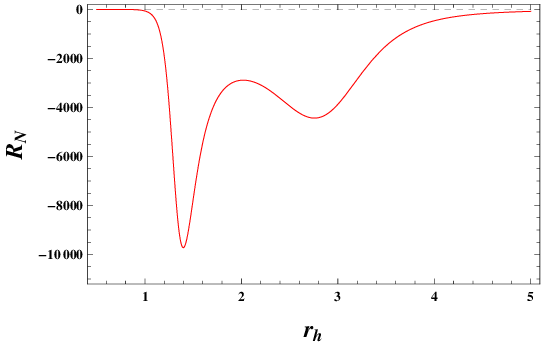}}}
\caption{$R_{\rm N}$ as a function of the horizon radius $r_{\text{h}}$ for different values of the temperature with $\alpha$=3.15. (a) $T$=0.062288. (b) $T$=0.0629. (c) $T$=$T_{\text{c1}}$=0.063320. (d) $T$=0.0634. (e) $T$=$T_{\text{c2}}$=0.063541. (f) $T$=0.064.}\label{ppRnrht064f}
\end{figure}
%%%%%%%%%%%%%%%%%%%%%%%%%%%

For the purpose that uncovering the detailed information of $R_{\text{N}}$, we show its behavior for different temperatures in Fig.~\ref{ppRnrht064f}. First, we set $T = 0.062288$, for which there are two negative divergent points and a small well appears near $r_{\text{h}}=2$, see Fig.~\ref{Rnrhtta}. For a lower temperature, the situation is similar. For $T=0.0629$, these two divergent points get a little closer. Meanwhile, the depth of the well increases, see Fig.~\ref{Rnrht0629b}. Further increasing the temperature such that it approaches the first critical point, i.e., $T=T_{\text{c1}}=0.063320$, we show the case in Fig.~\ref{Rnrhtc1c}. These two negative divergent points are still present. While the depth of the well approaches negative infinity. So there are three divergent points. When $T=0.0634$ described in Fig.~\ref{Rnrht0634d}, the behavior with third divergent points disappears and only a well with finite depth leaves. When the temperature approaches its second critical point (see Fig. \ref{Rnrhttc2e}), the first two divergent points coincide, and thus only one divergent point is present. Further increasing the temperature, all the divergent points disappear and there are only two wells, see Fig. \ref{Rnrht064f} for an example. These results show a general pattern of $R_{\text{N}}$ near the triple point.

Since the sign and divergent behavior of $R_{\text{N}}$ are very important for the Ruppeiner geometry, here we turn to examine them. Employing the expression (\ref{rrrn}) of $R_{\text{N}}$, we can easily obtain the temperatures $\tilde{T}_{0}$ and $\tilde{T}_{\text{sp}}$ corresponding to the sign-changing curve and spinodal curve, respectively,
\begin{eqnarray}
 \tilde{T}_{0}=\frac{1}{2}\tilde{T}_{\text{sp}}
 =\frac{3r_{\text{h}}^6 +2\alpha r_{\text{h}}^4-2}
       {4\pi r_{\text{h}}^5\left(r_{\text{h} }^2+6\alpha\right)},
\end{eqnarray}
along which the normalized curvature scalar vanishes or diverges. Moreover, another sign-changing curve can also be found by solving
\begin{eqnarray}
 3r_{\text{h}}^6+2\alpha r_{\text{h}}^4-2=0.
\end{eqnarray}
It is clear that this sign-changing curve is independent of the temperature. When $\alpha$=3.15, we have $r_{\text{h}}=0.7112$.

%%%%%%%%%%%%%%%%%%%%%%%%%%%
\begin{figure}
\center{\subfigure[]{\label{RNVIa}
\includegraphics[width=7cm]{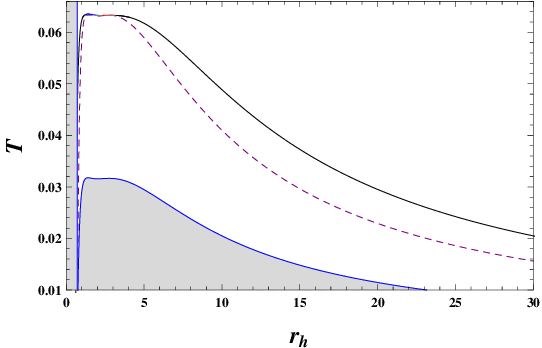}}
\subfigure[]{\label{RNVILarb}
\includegraphics[width=7cm]{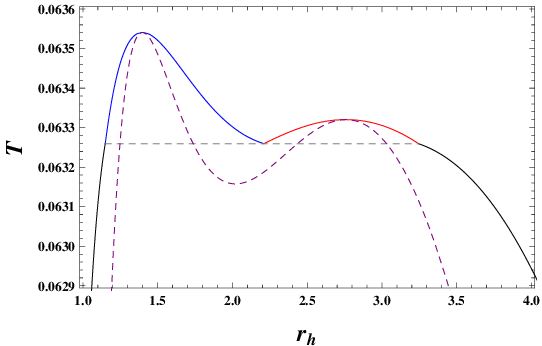}}}
\caption{Behavior of $R_{\text{N}}$ in the $T$-$r_{\text{h}}$ phase diagram for $\alpha$=3.15. $R_{\text{N}}$ is positive in the shadow regions, while negative in other regions. (a) The coexistence curve is described by the top solid curve. The purple dashed curve is for the spinodal curve. The left blue vertical line and the bottom blue curve are for the sign-changing curves. (b) Enlarged drawing of (a) near the triple point.}\label{ppRNVILarb}
\end{figure}
%%%%%%%%%%%%%%%%%%%%%%%%%%%

We show these temperatures in the $T$-$r_{\text{h}}$ phase diagram for $\alpha$=3.15 in Fig. \ref{ppRNVILarb}. The dashed curve is the spinodal curve, while the bottom blue curve and the left blue vertical line are for the sign-changing curves. In the shadow regions, $R_{\text{N}}$ is positive, which indicates that there is a repulsive interaction among the black hole microstructure according to the empirical observation of the Ruppeiner geometry. In other regions, $R_{\text{N}}$ is negative and attractive interaction dominates. Considering that in the coexistence region, we do not know whether the equation of state is applicable, we should exclude the region, which is because that we have used the equation of state (\ref{equat}) in obtaining the normalized curvature scalar. Following this idea, only the left shadow region is allowed. This result suggests that $R_{\text{N}}$ of the high temperature small black hole is positive and the repulsive interaction dominates among its microstructure. In other regions, only the attractive interaction dominates. This result is similar to that of other charged black holes in Einstein and GB gravities. In order to show the details near the triple point, we present an enlarged drawing in Fig. \ref{RNVILarb}. From it, we see that $R_{\text{N}}$ is always negative. The spinodal curve and the coexistence curve only coincide at the critical point, which implies that $R_{\text{N}}$ goes to negative infinity at the critical point. While at the triple point marked with the dashed horizontal line, $R_{\text{N}}$ maintain finite values.

\section{From critical point to triple point}
\label{fcpttp}

In order to compare our results with that of different fluids studied in Ref. \cite{Ruppeinerpre} by using the experimental data, in this section we examine the varying behaviors of the normalized curvature scalar $R_{\text{N}}$ along the coexistence intermediate-large black hole curve, reminiscent of the liquid-vapor coexistence curve.

At first, we briefly summarize the results obtained in Ref. \cite{Ruppeinerpre}, where the author clearly showed the curvature scalar along the liquid-vapor coexistence curve for six representative fluids, i.e., hydrogen, helium, argon, methane, oxygen, and water. One universal observation is that at the critical point, all the scalars have negative divergent behaviors. In particular, the critical exponent and universal parameter can be found following the treatment of Refs. \cite{Weiwa2,WeiWeiWei}. However, we will not calculate them here. For hydrogen and helium, the curvature scalars cross the $x$-axis at a certain temperature and become positive when one decreases the temperature from the critical point to the triple point along the coexistence liquid curve. However, along the coexistence vapor curve, the scalar is always negative. For the hydrogen, it slightly decreases at the triple point. Another different feature from other fluid is that the scalars of the coexistence liquid and vapor curves do not intersect. For argon and methane, the scalars have no zero point from the critical point to the triple point, while they intersect at some points. Along the coexistence liquid curve, the scalar increases from the critical point, and tends to zero at the triple point. While when along the coexistence vapor curve, these scalars get a slight decrease near the triple point. For oxygen and water, except the zero points, the significant difference is that the scalar of the coexistence vapor phase has large negative values near the triple points.

%%%%%%%%%%%%%%%%%%%%%%%%%%%
\begin{figure}
\center{\subfigure[]{\label{RNTcoex305a}
\includegraphics[width=7cm]{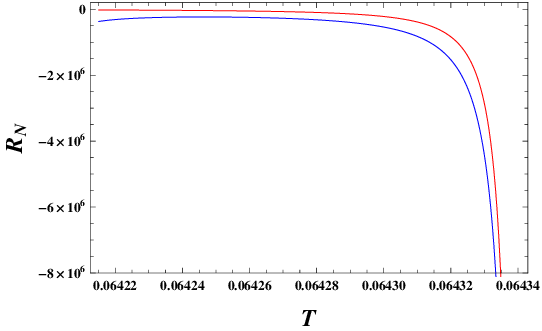}}
\subfigure[]{\label{RNTcoex315b}
\includegraphics[width=7cm]{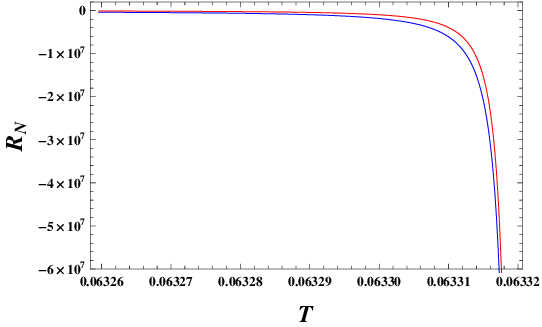}}}
\caption{The normalized curvature scalar $R_{\text{N}}$ along the coexisting intermediate and large black hole curves from the critical point to the triple point. The bottom blue and top red curves are for the coexisting intermediate and large black holes, respectively. (a) $\alpha$=3.05. (b) $\alpha$=3.15.}\label{ppRNTcoex315b}
\end{figure}
%%%%%%%%%%%%%%%%%%%%%%%%%%%

We exhibit $R_{\text{N}}$ along the coexisting intermediate and large black hole curves in Fig. \ref{ppRNTcoex315b}. Both for $\alpha$=3.05 and 3.15, we observe that $R_{\text{N}}$ negatively diverges at the critical points. For $\alpha$=3.05, see Fig. \ref{RNTcoex305a}, we find that $R_{\text{N}}$ gradually increases along the coexisting large black hole curve with the decrease of the temperature from the critical point to the triple point. However, along the coexisting intermediate black hole curve, it first increases and then slightly decreases near the triple point. This pattern is quite similar to argon and methane. When $\alpha$=3.15, the only difference is that the slightly decreased behavior disappears. Especially, there is no zero point of $R_{\text{N}}$. On the other hand, we expect that other patterns of $R_{\text{N}}$ could be discovered for other $\alpha$.

%%%%%%%%%%%%%%%%%%%%%%%%%%%
\begin{figure}
\center{\includegraphics[width=7cm]{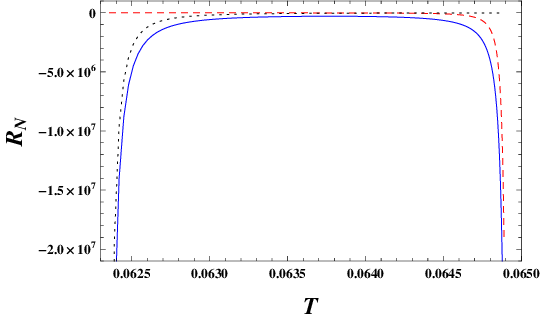}}
\caption{The normalized curvature scalar $R_{\text{N}}$ at the triple point. The red dashed curve, blue solid curve, and black dotted curve are, respectively, for the coexisting small, intermediate, and large black holes.}\label{RnT}
\end{figure}
%%%%%%%%%%%%%%%%%%%%%%%%%%%

Furthermore, we show $R_{\text{N}}$ at the triple point for the coexisting small, intermediate, and large black holes in Fig. \ref{RnT}. Interestingly, we observe two divergent behaviors at the lowest and highest temperatures of the triple points. This is mainly because that the triple point coincides with the critical point at the lowest and highest temperatures. If the strength of the interaction is characterized by the value of $R_{\text{N}}$, this suggests in the middle temperature, the interactions of these three components compete with each other.

In summary, after considering the normalized curvature scalar from the critical point to the triple point, the results suggest that the six-dimensional charged GB-AdS black hole is more like argon or methane.

\section{Conclusions and discussions}
\label{Conclusion}

In this paper, we have studied the Ruppeiner geometry and phase transition for the six-dimensional charged GB-AdS black holes characterized by the triple point.

We performed a detailed study of the phase transition near the triple point, and several typical features were uncovered. The swallow tail behavior is clearly exhibited. Near the triple point, a double swallow tail behavior is present. Especially, at the triple point, the intersection points coincide for the two-swallow tail behavior, which indicates a coexistence point of the three different black hole phases.

In addition, the equal area law is applicable. Below the triple point, a conventional pattern of the equal area law was observed, where two equal areas were constructed by one isothermal curve and a horizontal line with constant pressure. At the triple point, two pairs of equal areas were observed. Similarly, these four areas were constructed by one isothermal curve and one horizontal line, which indicates these two phase transitions have the same temperature and pressure. Slightly above the triple point, two pairs of equal areas were also found. However, they were constructed by two different horizontal lines. Further increasing the temperature, a conventional equal area law holds. Above the top critical point, no phase transition exists, and thus no equal areas can be constructed any more.

Based on the detailed analysis, we clearly exhibited the phase structures both in the $P$-$T$ and $T$-$r_{\text{h}}$ diagrams. The small, intermediate, and large black hole phases, as well as the triple points and critical points were shown. Interestingly, the degeneracy of the coexistence curve in the $P$-$T$ phase diagram could be broken, and the coexistence regions were explicitly shown in the $T$-$r_{\text{h}}$ phase diagram. Of particular interest is in the $T$-$r_{\text{h}}$ diagram, the phase structure is much like a double hill pattern, while with different heights.

Then we constructed the Ruppeiner geometry for the black hole. By making use of the equation of state, we calculated the corresponding normalized curvature scalar. Different from the small-large black hole phase transition of the VdW-type, we found that the scalar has a richer behavior near the triple point. Especially, more than two negatively divergent points were observed. Furthermore, via the normalized curvature scalar, the spinodal curve and the sign-changing curve were obtained. And they are presented in the $T$-$r_{\text{h}}$ phase diagram. When excluding the coexistence region, we observed that the normalized curvature scalar is positive for the high temperature small black hole, which implies that the repulsive interaction dominates among these black hole microstructure. This property is similar to that of the other charged or rotating AdS black hole systems. On the other hand, near the triple point, the scalar always keeps a finite negative value except the critical point, where the spinodal curve coincides with the coexistence curve and the scalar goes to negative infinity.

Varying the temperature from the triple point to the critical point, we calculated the normalized curvature scalar along the coexisting intermediate and large black hole curves, reminiscent of the liquid-vapor coexistence curve. When $\alpha$=3.05, $R_{\text{N}}$ increases along the coexisting large black hole curve from the critical point to the triple point. While along the coexisting intermediate black hole, $R_{\text{N}}$ firstly increases and then slightly decreases near the triple point. This behavior is similar to argon or methane. When $\alpha$=3.15, the pattern almost keeps the same except that the slightly decrease disappears near the triple point.

To summarize, we studied the thermodynamics and Ruppeiner geometry for the charged GB-AdS black hole near the triple point. On one hand, we disclosed the general feature of the thermodynamics near the triple point including the multi-equal area law, the swallow tail behaviors and the phase structure. On the other hand, via the normalized curvature scalar, it suggests that the black hole behaves more like argon or methane. These uncover the particular novel microscopic properties for the charged AdS black holes near the triple point in six-dimensional GB gravity. The results are also worth to extend to other black hole system with more richer phase structure including the superfluid phase and isolated critical point.

\section*{Acknowledgements}
The authors would like to thank Prof. George Ruppeiner for his useful discussions. This work was supported by the National Natural Science Foundation of China (Grants No. 12075103, No. 11675064, No. 11875151, and No. 12047501), the 111 Project (Grant No. B20063), and the Fundamental Research Funds for the Central Universities (No. Lzujbky-2019-ct06).

\end{document}